\documentclass[aps,onecolumn,showpacs]{revtex4}
\usepackage{graphicx}
\usepackage{color}
\usepackage{natbib}
\newcommand{\tbox}[1]{\mbox{\tiny #1}}

\begin{document}

\title{Non-uniform random graphs on the plane: A scaling study}

\author{C. T. Mart{\'\i}nez-Mart{\'\i}nez,$^1$ J. A. M\'endez-Berm\'udez,$^1$ Francisco A. Rodrigues,$^2$ and Ernesto Estrada$^3$}

\affiliation{
$^1$Instituto de F\'{\i}sica, Benem\'erita Universidad Aut\'onoma de Puebla,
Apartado Postal J-48, Puebla 72570, Mexico \\
$^2$Departamento de Matem\'{a}tica Aplicada e Estat\'{i}stica, Instituto de Ci\^{e}ncias 
Matem\'{a}ticas e de Computa\c{c}\~{a}o, Universidade de S\~{a}o Paulo - Campus de S\~{a}o Carlos, 
Caixa Postal 668, 13560-970 S\~{a}o Carlos, SP, Brazil \\
$^3$Institute for Cross-Disciplinary Physics and Complex Systems (IFISC-CSIC-UIB),
Campus Universitat de les Illes Balears E-07122, Palma de Mallorca, Spain}

\begin{abstract}
We consider random geometric graphs on the plane characterized by a 
non-uniform density of vertices.
In particular, we introduce a graph model where $n$ vertices are independently distributed in the unit 
disc with positions, in polar coordinates $(l,\theta)$, obeying the probability density functions $\rho(l)$
and $\rho(\theta)$.
Here we choose $\rho(l)$ as a normal distribution with zero mean and variance $\sigma\in(0,\infty)$
and $\rho(\theta)$ as an uniform distribution in the interval $\theta\in [0,2\pi)$.
Then, two vertices are connected by an edge if their Euclidian distance is less or equal than the 
connection radius $\ell$.
We characterize the topological properties of this random graph model, 
which depends on the parameter set $(n,\sigma,\ell)$, by the use of the average
degree $\left\langle k \right\rangle$ and the number of non-isolated vertices $V_\times$; while we approach their 
spectral properties with two 
measures on the graph adjacency matrix: the ratio of consecutive eigenvalue spacings $r$ and the 
Shannon entropy $S$ of eigenvectors.
First we propose a heuristic expression for $\left\langle k(n,\sigma,\ell) \right\rangle$.
Then, we look for the scaling properties of the normalized average measure $\left\langle \overline{X} \right\rangle$ 
(where $X$ stands for $V_\times$, $r$ and $S$) over graph ensembles.
We demonstrate that the scaling parameter of $\left\langle \overline{V_\times} \right\rangle=\left\langle V_\times \right\rangle/n$ 
is indeed $\left\langle k \right\rangle$; with $\left\langle \overline{V_\times} \right\rangle \approx 1-\exp(-\left\langle k \right\rangle)$.
Meanwhile, the scaling parameter of both $\left\langle \overline{r} \right\rangle$ and $\left\langle \overline{S} \right\rangle$ is proportional 
to $n^{-\gamma} \left\langle k \right\rangle$ with $\gamma\approx 0.16$.
\end{abstract}
\maketitle

\date{\today}



\section{Introduction}

In many complex systems the entities are embedded in a geometric space
such as the connections between them are mainly determined by their
spatial separations. When representing these systems we deal with
the so-called spatial networks~\cite{Spatial networks}, where nodes
representing the entities of the system are located in a given space,
which may include a geographic space, like in the case of road networks~\cite{Spatial networks}. 
Nodes can also be located in a three-dimensional space like in certain
biological networks where the proximity of cells in a biological tissue
determine the structure of the network (see~\cite{EstradaBook}).
The list of related examples includes the networks of patches and corridors
in a landscape~\cite{Landscape networks}, the networks of galleries
in animal nests~\cite{Termites,Ants}, and the networks of fractures
in rocks~\cite{Rock fractures}, networks of wireless communication
devices~\cite{RGG wireless,RGG communication,RGG comm}, such as mobile
phones, wireless computing systems, wireless sensor networks, among
others.

In modeling these systems most of the efforts are a continuation of
the pioneering work of Gilbert in 1959 when he proposed the randon
geometric graph (RGG) model~\cite{Gilbert model}. In a RGG the nodes
of the graph are distributed randomly and independently in a unit
square and two nodes are connected if they are inside a disk of a
given radius~\cite{Penrose,Dall Christense}. This kind of random
graphs have found multiple applications in areas such as modelling
of epidemic spreading in spatial populations, which may include cases
such the spreading of worms in a computer network, viruses in a human
population, or rumors in a social 
network~\cite{Spatial connectivity,RGG Sync,Worm Epidemics,RGG spreading,RPG epidemics}.
Recent modifications known as random rectangular graphs (RRG) have
been applied in a variety of physical scenarios~\cite{RRG_1,RGG_2,RGG_3,RGG_4}.

Gilbert model and their analogous are very useful in situations where
the nodes are uniformly distributed in the graph. This is the case
for instance when we want to deploy a series of wireless sensors in
a given area. In this case RGG or RRG are very useful modeling techniques
to characterize the properties of the network emerging from the sensor
deployment~\cite{WSN}. However, in the scenarios mentioned before
where the nodes are embedded in a given space not necessarily in a
uniform way, the consideration of RGG/RRG is not the most appropriate
one.

Here we propose an extension of the RGG model to consider different
degrees of spatial non-uniformity in the graph. First, we characterize the average 
topological properties of this random graph model.
In particular, we propose heuristic expressions for the average degree  
and the number of non-isolated vertices. 
Then, within a random matrix theory 
approach, we characterize the spectral and eigenvector properties of the adjacency matrix.
To this end we perform a scaling study of both, the ratio of consecutive eigenvalue spacings 
and the Shannon entropy of eigenvectors.

\section{Preliminaries}

Let $G=\left(V,E\right)$ be a graph where $V$ is the set of vertices and $E$ the set of edges. 
A RGG is the undirected graph with $n=|V|$ randomly sampled vertices in $[0,1)^2$, where two 
vertices are connected by an edge if their euclidian distance is less or equal than the parameter
$\ell$. Here, $\ell$ is known as the connection radius.

We consider propose an extension of the RGG model where $n$ vertices are independently 
distributed in the unit disc with 
positions, in polar coordinates $(l,\theta)$, obeying the probability density functions $\rho(l)$
and $\rho(\theta)$, respectively. Here we choose $\rho(l)$ as a normal distribution with zero mean 
and variance $\sigma$, $\rho(l) = {\cal N}(0,\sigma^2)$, and $\rho(\theta)$ as an uniform distribution 
in the interval $\theta\in (0,2\pi)$, $\rho(\theta) = {\cal U}(0,2\pi)$.
Indeed, the parameter $\sigma\in(0,\infty)$ accounts for the degree of spatial non-uniformity of the 
graph; for $\sigma<1$ a cluster of vertices is formed around the disc center, while for $\sigma \gg 1$
the distribution of vertices becomes uniform within the unit disc.
Then, two vertices are connected by an edge if their euclidian distance is less or equal than the 
connection radius $\ell\in[0,2]$; where 2 corresponds to the diameter of the unit disc, so it is the
maximum value that $\ell$ can take.
Therefore, this random graph model depends on three parameters: the number of vertices $n$, the 
degree of non-uniformity $\sigma$ and the connection radius $\ell$.
Note that both $\sigma$ and $\ell$ are given in units of the disc radius, chosen here to be one.

\begin{figure*}[t!]
\begin{center} 
\includegraphics[width=0.65\textwidth]{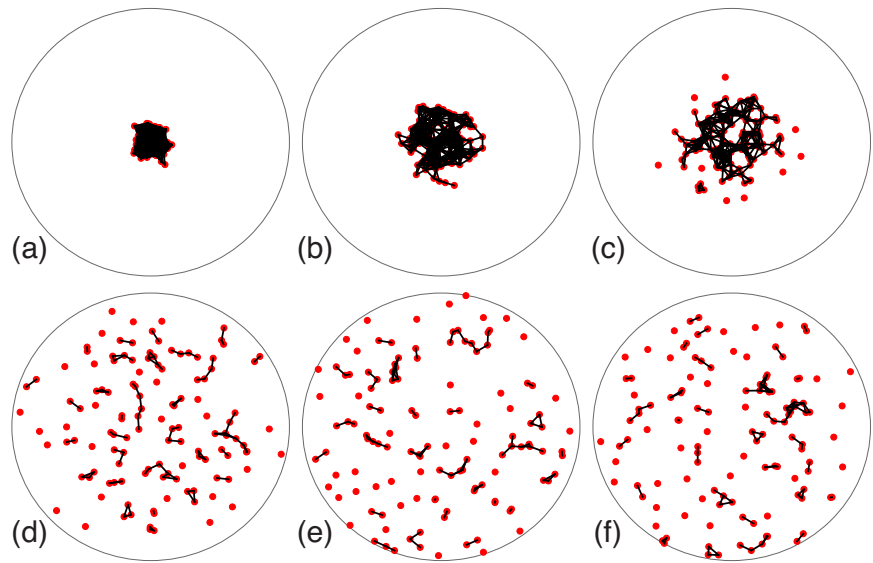}
\caption{
Examples of RGGs with different degrees of non-uniformity $\sigma$: 
(a) $\sigma = 0.01$, (b) $\sigma = 0.04$, (c) $\sigma = 0.1$, (d) $\sigma = 0.4$, 
(e) $\sigma = 1$ and (f) $\sigma = \infty$.
In all cases we consider $n=125$ vertices and a connection radius of $\ell=0.1$.}
\label{Fig01}
\end{center}
\end{figure*}

In Fig.~\ref{Fig01} we present examples of RGGs with different degrees of 
non-uniformity $\sigma$ along several orders of magnitude. In all cases we consider $n=125$ 
vertices and a connection radius of $\ell =0.1$. 
As can be clearly seen from Fig.~\ref{Fig01}, our random graph model produces a cluster around 
the disc center for $\sigma<1$. Formally, it reproduces the RGG model (in the disc) 
when $\sigma\to\infty$, however, as can be observed in Fig.~\ref{Fig01}(d), already for $\sigma=1$
the vertices appear uniformly distributed. In fact, as will be shown below, 
$\sigma_c\approx 1$ separates two graph regimes: {\it the clustering regime} when 
$\sigma<\sigma_c$ and {\it the uniform regime} when $\sigma\ge \sigma_c$.
We want to add that we choose a disc as the embedding geometry, instead of a square as in
other studies of RGGs, to account for the radial symmetry of the vertex distribution.

We characterize the topological properties of our random graph model by the use of the average
degree $\left\langle k \right\rangle$ and the number of non-isolated vertices $V_\times$. While we approach their 
spectral properties with two well-known RMT measures on the graph adjacency matrix: the ratio of 
consecutive eigenvalue spacings $r$ and the Shannon entropy $S$ of eigenvectors.

On the one hand, it is relevant to stress that analytical expressions for both $\left\langle k \right\rangle$ and 
$\left< V_\times \right>$ on RGGs have been reported recently. In fact, we will make use of those 
expressions to approach our model of non-uniform RGGs.
In particular:

(i) In Ref.~\cite{RRG_1} the expression for the average degree of RGGs embedded in the 
unit rectangle was derived; it redas as 
\begin{equation}
\label{k}
\left< k \right> = (n-1)f \, ,
\end{equation}
where $f$ is a highly nontrivial function of the connection radius $\ell$ and the rectangle side lengths.
Clearly, when the rectangle sides are equal the expression for $\left< k \right>$ of RGGs embedded 
in the unit square is obtained; in such a case $f$ gets the form
\begin{equation}
f(\ell) = 
\left\{ 
\begin{array}{ll}
           \ell^2  \left[ \pi - \frac{8}{3}\ell +\frac{1}{2}\ell^2 \right] & \quad 0 \leq \ell \leq 1 \, , 
           \vspace{0.25cm} \\
             \frac{1}{3} - 2\ell^2 \left[ 1 - \arcsin(1/\ell) + \arccos(1/\ell) \right] 
             +\frac{4}{3}(2\ell^2+1) \sqrt{\ell^2-1} 
             -\frac{1}{2}\ell^4 & \quad 1 \leq \ell \leq \sqrt{2} \, .
\end{array}
\right.
\label{f(ell)}
\end{equation}

(ii) The average of $V_\times$ can be computed from the average number 
of isolated vertices $\left< K_1 \right>$ as $\left< V_\times \right>=n-\left< K_1 \right>$.
In fact, for standard RGGs, $\left< K_1 \right>$ is already known~\cite{DMP07}; it is given by
$$
\left< K_1 \right>=n (1-\pi \ell^2)^{n-1}=n\exp(-n\pi \ell^2)-O(n\ell^4) \, .
$$
Therefore, for standard RGGs, we can write
\begin{equation}
\left< V_\times \right> \approx n\left[1-\exp(-n \pi \ell^2)\right] \, .
\label{Vxofell}
\end{equation}

On the other hand, given the ordered spectra $\{ \lambda_i \}$ ($i=1,\ldots, n$) and the corresponding 
normalized eigenvectors $\Psi^i$ (i.e.~$\sum_{j=1}^n | \Psi^i_j |^2 =1$) of an adjacency matrix, the 
$i$-th \emph{ratio of consecutive eigenvalue spacings} 
is given by~\cite{ABG13}
\begin{equation}
\label{r}
r_i = \frac{\min(\lambda_{i+1}- \lambda_i,\lambda_{i}- \lambda_{i-1})}{\max(\lambda_{i+1}- \lambda_i,\lambda_{i}- \lambda_{i-1})} \ ,
\end{equation}
while the \emph{Shannon entropy} of the eigenvector $\Psi^i$ reads as
\begin{equation}
\label{S}
S_i = -\sum_{j=1}^n \left| \Psi^i_j \right|^2 \ln \left| \Psi^i_j \right| ^2 \ .
\end{equation}
We would like to mention that in contrast to the Shannon entropy which is a well accepted 
quantity to measure the degree of disorder in complex networks, the use of the ratio of consecutive 
eigenvalue spacings is relative recent in graph studies; see for example~\cite{AMRS20,TFM20,MAMPS19,PRR20}.

Here, we will follow a recently introduced approach under which the adjacency matrices of random 
graphs are represented by RMT ensembles; see the application of this approach on Erd\"os-R\'{e}nyi 
graphs~\cite{TFM20,MAM15}, RGGs and random rectangular graphs~\cite{AMGM18}, 
$\beta$-skeleton graphs~\cite{AME19}, multiplex and multilayer networks~\cite{MFMR17}, and 
bipartite graphs~\cite{MAMPS19}. Consequently, we define the elements of the adjacency matrix 
$\mathbf{A}$ of our random graph model as
\begin{equation}
A_{uv}=\left\{
\begin{array}{cl}
\sqrt{2} \epsilon_{uu} \ & \mbox{for $u=v$}, \\
\epsilon_{uv} & \mbox{if there is an edge between vertices $u$ and $v$},\\
0 \ & \mbox{otherwise}.
\end{array}
\right.
\label{A}
\end{equation}

Here, we choose $\epsilon_{uv}$ as statistically-independent random variables drawn from a normal 
distribution with zero mean and variance one. Also, $\epsilon_{uv}=\epsilon_{vu}$, since our graphs are 
assumed as undirected. 
According to this definition, diagonal random matrices are obtained for $\ell=0$ (Poisson ensemble 
(PE), in RMT terms), whereas the Gaussian Orthogonal Ensemble (GOE) (i.e.~full real and symmetric 
random matrices) is recovered when $\ell=2$. Therefore, a transition from the PE to the GOE can be 
observed by increasing $\ell$ from zero to two, for any given fixed pair $(n,\sigma)$. 
In fact, this is not the only way to observe the PE to GOE transition; it could also be observed
by decreasing $\sigma$ for a given fixed pair $(n,\ell)$, see e.g.~Fig.~\ref{Fig01}, or by increasing
$n$ for fixed $(\sigma,\ell)$.

Notice that the random weights we impose to the adjacency matrix in~(\ref{A}) do not play any role in 
the computation of $\left< k \right>$ nor $\left< V_\times \right>$, however these weights help us obtaining 
non-null adjacency matrices (that 
we can still diagonalize) for graphs with a large number of isolated vertices; so we can safely explore 
numerically the spectral and eigenvector properties of the model in the limit $\ell\to 0$. 

From the definitions above, when $\ell=0$ (i.e.~when all vertices in the graph are isolated) we have 
$\left< k \right>_{\tbox{PE}}=0$, $\left< V_\times \right>_{\tbox{PE}}=0$, 
$\left< r \right>_{\tbox{PE}}\approx 0.3863$~\cite{ABG13}, and $\left< S \right>_{\tbox{PE}}=0$.
While when $\ell=2$ (i.e.~when the graph is complete), $\left< k \right>_{\tbox{GOE}}=n-1$,
$\left< V_\times \right>_{\tbox{GOE}}=n$, $\left< r \right>_{\tbox{GOE}}\approx 0.5359$~\cite{ABG13}, 
and $\left< S \right>_{\tbox{GOE}}\approx \ln (n/2.07)$~\cite{MK98}.
Here and below $\left< \cdot \right>$ denotes the average over an ensemble of adjacency matrices 
$\mathbf{A}$, in the case of $k$ and $V_\times$, and the average over all eigenvalues [eigenvectors] 
of an ensemble of adjacency matrices $\mathbf{A}$ in the case of $r$ [$S$].
We want to add that the predictions for $\left< r \right>_{\tbox{PE}}$, $\left< r \right>_{\tbox{GOE}}$ 
and $\left< S \right>_{\tbox{GOE}}$ reported above are expected for large $n$; i.e.~finite size 
effects should be observed for small $n$, typically for $n<100$ (see e.g.~\cite{PRR20}).

\begin{figure*}[t!]
\begin{center} 
\includegraphics[width=0.75\textwidth]{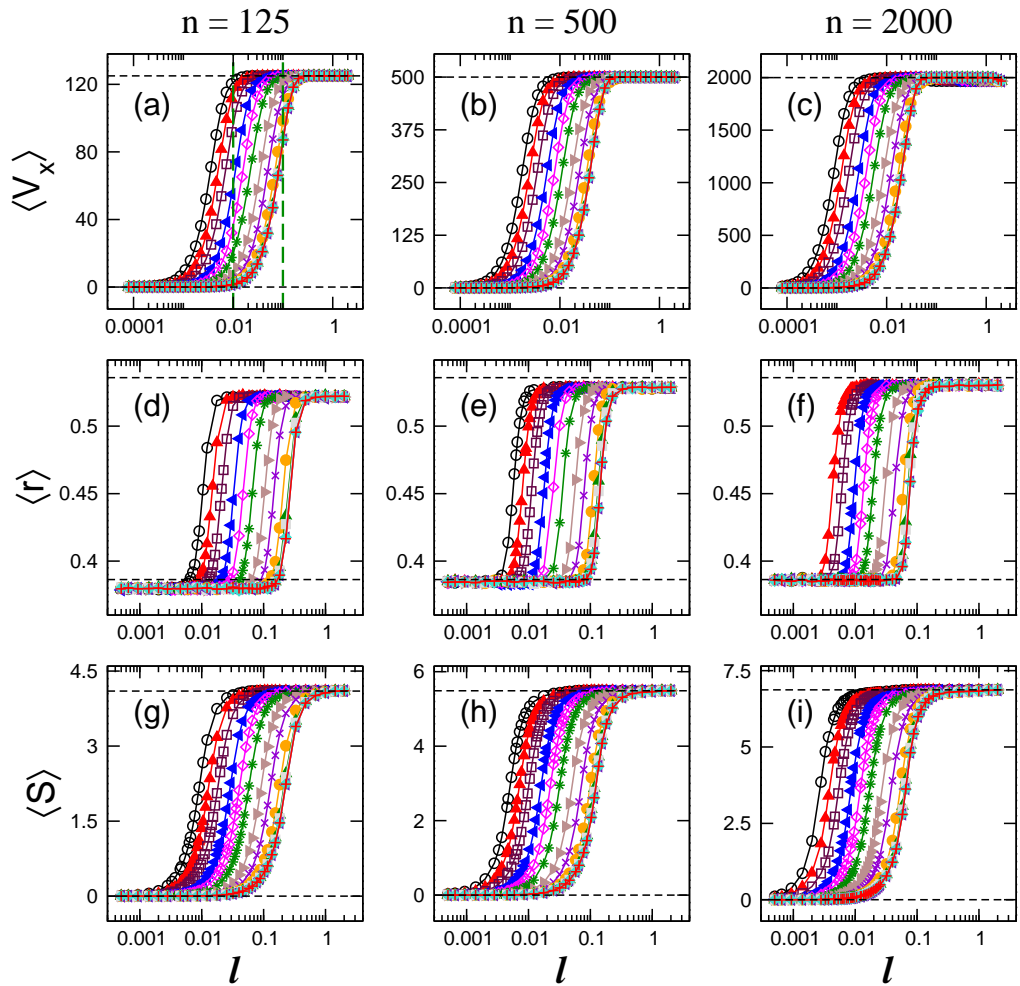}
\caption{
(a-c) Average number of non-isolated vertices $\left\langle V_\times \right\rangle$,
(d-f) average ratio of consecutive eigenvalue spacings $\left< r \right>$, and
(g-i) average Shannon entropy $\left< S \right>$
as a function of the connection radius $\ell$ of non-uniform RGGs of size $n$: 
(a,d,g) $n=125$, (b,e,h) $n=500$, and (c,f,i) $n=2000$.
Each panel displays 14 curves corresponding to different degrees of non-uniformity 
$\sigma$: $\{0.001, 0.002, 0.004, 0.01, 0.02, 0.04, 0.1, 0.2, 0.4, 0.8, 1, 10, 100, \infty \}$, 
form left to right.
Horizontal dashed lines (lower and upper, respectively) indicate the predictions for the PE and GOE limits.
The vertical dashed lines in (a) mark $\ell=0.01$ and 0.1; 0.1 is the value of $\ell$ used in Fig.~\ref{Fig01}.
Each data value was computed by averaging over $10^7/n$ random graphs.
}
\label{Fig02}
\end{center}
\end{figure*}

\section{Computation of average measures}

Now, in Fig.~\ref{Fig02} we present $\left< V_\times \right>$, $\left< r \right>$, and $\left< S \right>$ 
as a function of the connection radius $\ell$ of non-uniform RGGs of size $n$ (we will consider 
$\left< k \right>$ later on).
In this figure we are using three values of $n$: $n=125$ (left panels), $n=500$ (middle panels), 
and $n=2000$ (right panels). Each panel displays 14 curves corresponding to different degrees 
of non-uniformity $\sigma$ (increasing form left to right).
In this figure we can clearly see the effect of the parameter $\sigma$ on the properties of the
random graph model: For fixed graph size and fixed connection radius, see for example the vertical
dashed line in Fig.~\ref{Fig02}(a) at $\ell=0.01$, the graphs may transit from mostly connected (see the 
left-most curve corresponding to $\sigma=0.001$) to mostly disconnected (see the right-most curve 
corresponding to $\sigma=\infty$). This panorama was already shown in Fig.~\ref{Fig01}, however in 
that example the graph does not become disconnected even when $\sigma\to\infty$ due to the
use of a larger value of $\ell$: $\ell=0.1$, see the vertical dashed line in Fig.~\ref{Fig02}(a).

Indeed, several facts can be highlighted from Fig.~\ref{Fig02}:
\begin{itemize}
\item[(i)] All curves $\left< X \right>$ vs.~$\ell$ in all panels show a smooth transition (in 
semi-log scale) from the PE to the GOE for increasing $\ell$; the PE and GOE limits are 
indicated as horizontal dashed lines (lower and upper, respectively). 
Here and below $X$ stands for $V_\times$, $r$ or $S$.
\item[(ii)] For fixed $n$ [$\sigma$] the curves $\left< X \right>$ vs.~$\ell$ have a very similar 
functional form but they are displaced to the right on the $\ell$-axis for increasing $\sigma$ [$n$], 
thus
\item[(iii)] the onset of the GOE limit is reached for smaller values of $\ell$ the larger the values 
of $n$ and $\sigma$ are, however
\item[(iv)] once $\sigma\ge \sigma_c$, with $\sigma_c\approx 1$, the curves $\left< X \right>$ vs.~$\ell$ 
do not change by further increasing $\sigma$.
\item[(v)] Small-size effects in $\left< r \right>$ are particularly 
visible when $\ell\to 0$ and $\ell\to 2$ in the case of $n=125$, see panel (d).
\end{itemize}

It is relevant to stress that we validated observation (iv) for many other graph sizes, which allowed
us to conclude that $\sigma_c \approx 1$ separates two regimes of our random graph model: {\it the 
clustering regime} when $\sigma<\sigma_c$ and {\it the uniform regime} when $\sigma\ge \sigma_c$.
In the clustering regime we do observe a cluster of vertices around the disc center, see e.g.~Figs.~\ref{Fig01}(a-c),
while in the uniform regime the vertices are evenly distributed over the unit disc, see e.g.~Figs.~\ref{Fig01}(d,e).

Moreover, given the similar functional form of the curves $\left< X \right>$ vs.~$\ell$ for different combinations 
of $n$ and $\sigma$, as reported in Fig.~\ref{Fig02}, it seems that they could be effectively scaled. 
That is, one should be able to find a scaling parameter $\xi\equiv\xi(n,\sigma,\ell)$ such that the 
curves $\left< \overline{X} \right>$ vs.~$\xi$ are invariant, where $\overline{X}$ is the properly 
normalized measure $X$. 

Since in previous studies of RGGs (embedded in the unit square) the average 
degree $\left< k \right>$ was shown to play a prominent role in the scaling of topological  
properties~\cite{AHMS20,AMRS20}, before performing the scaling analysis of our random graph model, in the 
next section we focus on its average degree.

\section{Average degree}

We numerically found that the expression for $\left< k \right>$ reported in Ref.~\cite{RRG_1}, for the 
particular case of RGGs embedded in the unit square, works pretty 
well for our model of non-uniform RGGs in the unit circle by properly choosing an effective connection 
radius $L$. That is, we propose the following heuristic expression for $f(L)$:
\begin{equation}
f(L) = 
\left\{ 
\begin{array}{ll}
           L^2  \left[ \pi - \frac{8}{3}L +\frac{1}{2}L^2 \right] & 0 \leq L \leq 1 \, , 
           \vspace{0.25cm} \\
             \frac{1}{3} - 2L^2 \left[ 1 - \arcsin(1/L) + \arccos(1/L) \right] 
              +\frac{4}{3}(2L^2+1) \sqrt{L^2-1} 
             -\frac{1}{2}L^4 & 1 \leq L \leq \sqrt{2} \, , 
             \vspace{0.25cm} \\
	     1 & L \geq \sqrt{2} \, .
\end{array}
\right.
\label{f(L)}
\end{equation}
When setting $L=\ell$, Eqs.~(\ref{k},\ref{f(L)}) provide the average degree of RGGs embedded in 
the unit square, see Eq.~(\ref{f(ell)}). Notice however that the last condition in Eq.~(\ref{f(L)}) was 
included to fit our 
random graph model; it does not apply to RGGs in the unit square since there $\ell$ cannot be larger 
than $\sqrt{2}$.

\begin{figure*}[t!]
\begin{center} 
\includegraphics[width=0.65\textwidth]{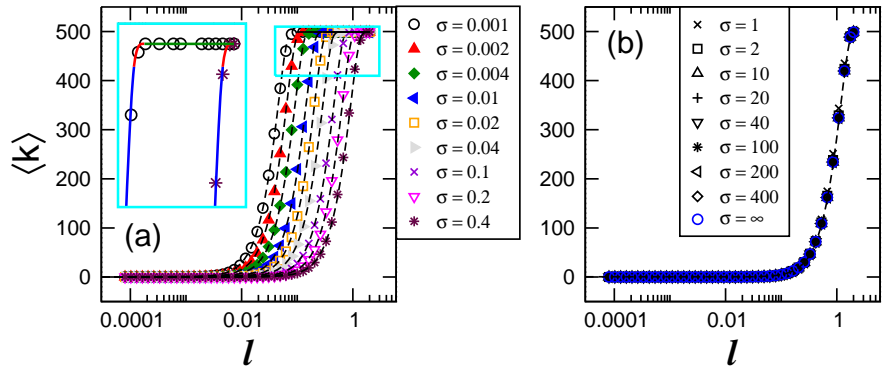}
\caption{
Average degree $\left< k \right>$ as a function of the connection radius $\ell$ of non-uniform RGGs of
size $N=500$ with (a) $\sigma < 1$ and (b) $\sigma \geq 1$.
Each data value was computed by averaging over $10^7/n$ random graphs.
Dashed lines correspond to Eqs.~(\ref{k},\ref{f(L)}). 
The inset in (a) is an enlargement of the cyan rectangle where the data for $\sigma = 0.001$ (circles) 
and $\sigma = 0.4$ (asterisks) is shown together with Eqs.~(\ref{k},\ref{f(L)}); there, blue, red and green 
lines correspond to the first, second and third conditions of Eq.~(\ref{f(L)}), respectively.
}
\label{Fig03}
\end{center}
\end{figure*}

Specifically, Eqs.~(\ref{k},\ref{f(L)}) provide a good approximation of $\left< k \right>$ of non-uniform RGGs 
in the unit circle if $L=\alpha\ell/\sqrt{\pi}$, with $\alpha=\sqrt{2/3\sigma}$ for $\sigma<\sigma_c$
and $\alpha=1$ for $\sigma\ge \sigma_c$; with $\sigma_c\approx 1$. 
To verify this claim, in Fig.~\ref{Fig03} we plot $\left< k \right>$ 
as a function of $\ell$ of non-uniform RGGs in the unit circle with both 
$\sigma < \sigma_c$ and $\sigma \geq \sigma_c$; there, the good correspondence between numerical 
calculations (symbols) and Eqs.~(\ref{k},\ref{f(L)}) (dashed lines) is evident. 
Moreover, in the inset of Fig.~\ref{Fig03}(a) we show the contribution of the three conditions of Eq.~(\ref{f(L)}) 
to the curves $\left< k \right>$ vs.~$\ell$ of two examples of non-uniform RGGs.
In Fig.~\ref{Fig03} we consider the fixed graph size $N=500$ but we observed
equivalent plots for any other graph size we tested.
 
We want to recall that $\sigma_c \approx 1$ indeed separates two regimes in the non-uniform RGG
model we study here: {\it the clustering regime} when 
$\sigma<\sigma_c$ and {\it the uniform regime} when $\sigma\ge \sigma_c$.
However, these two regimes will not be evident in the scaling we will perform below since, through the 
effective connection radius $L$, both are 
already incorporated in the definition of $\left< k \right>$ (that we will use to find the scaling parameters
of $\left< V_\times \right>$, $\left< r \right>$ and $\left< S \right>$).
Moreover, the two regimes can be clearly identified in a {\it straightforward} scaling analysis, as shown in
Subsection~\ref{Sub:straight}, see e.g.~Figs.~\ref{Fig12}(d-g).

\section{Scaling analysis}

\subsection{Average number of non-isolated vertices}
\label{Sub:Vx}

\begin{figure*}[t!]
\begin{center} 
\includegraphics[width=0.6\textwidth]{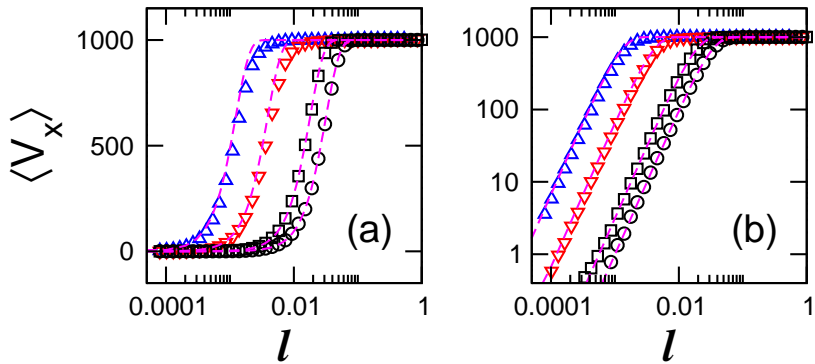}
\caption{
(a) Average number of non-isolated vertices $\left\langle V_\times \right\rangle$ as a function of the connection 
radius $\ell$ for RGGs embedded in the unit circle (circles), RGGs embedded in the unit square 
(squares) and non-uniform RGGs with $\sigma =0.01$ (triangles down) and $\sigma =0.001$ 
(triangles up); from right to left. In all cases $n=1000$ was used. 
Dashed lines correspond to Eq.~(\ref{Vx}).
(b) Same as in (a) but in log-log scale to better observe the region $\ell\ll 1$.
}
\label{Fig04}
\end{center}
\end{figure*}

Remarkably, taking as a reference Eq.~(\ref{Vxofell}), we found that 
\begin{equation}
\left< V_\times \right> \approx n\left[1-\exp(-n \pi L^2)\right] 
\label{Vx}
\end{equation}
approximates well
$\left< V_\times \right>$ of non-uniform RGGs in the unit circle if $L=\alpha\ell/\sqrt{\pi}$, with 
$\alpha=\sqrt{2/3\sigma}$ for $\sigma<\sigma_c$ and $\alpha=1$ for $\sigma\ge \sigma_c$;
this, in line with the proposal of Eq.~(\ref{f(L)}) from Eq.~(\ref{f(ell)}).
Thus, in Fig.~\ref{Fig04} we contrast Eq.~(\ref{Vx}) with numerical data. 
There we plot $\left< V_\times \right>$ as a function of $\ell$ for RGGs in the unit square, in the unit circle 
as well as for two examples of non-uniform RGGs. In all cases we observe good correspondence between 
numerical calculations (symbols) and Eq.~(\ref{Vx}) (dashed lines).

\begin{figure*}[t!]
\begin{center} 
\includegraphics[width=0.85\textwidth]{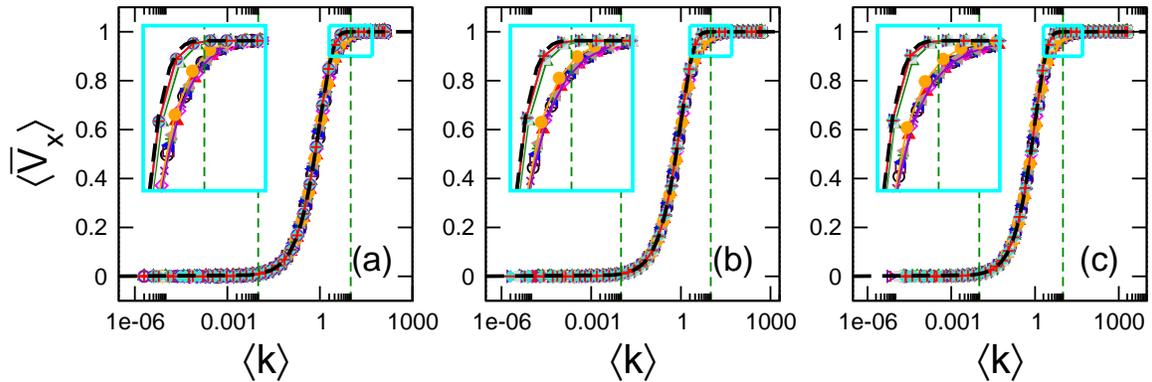}
\caption{
Average number of non-isolated vertices $\left\langle V_\times \right\rangle$, normalized to $n$, as a function of the 
average degree $\left< k \right>$ of non-uniform RGGs. (a) $n=125$, (b) $n=500$ and (c) $n=2000$.
Each panel displays the 14 curves reported in Fig.~\ref{Fig02}(a-c). 
The insets are enlargements of the cyan rectangles of the main panels. 
Dashed lines correspond to Eq.~(\ref{Vxofk}). Vertical dashed lines at $\left< k \right>=0.01$ and 10 
mark, approximately, the onset of delocalization and the onset of the GOE regime, respectively.
}
\label{Fig05}
\end{center}
\end{figure*}

Moreover, it is interesting to notice that from Eqs.~(\ref{k},\ref{f(L)}), when $L\ll 1$, we can write
$\left< k \right> \approx n\pi L^2$ which coincides with the argument of the exponential in~(\ref{Vx}).
This allows us to relate $\left< V_\times \right>$ and $\left< k \right>$ as 
\begin{equation}
\left< V_\times \right> \approx n\left[1-\exp(-\left< k \right>)\right] \, .
\label{Vxofk}
\end{equation}
Equation~(\ref{Vxofk}) implies that the scaling parameter, $\xi\equiv\xi(n,\sigma,\ell)$, of 
$\left\langle \overline{V_\times} \right\rangle=\left< V_\times \right>/n$ of non-uniform RGGs is in fact $\left< k \right>$; 
that is, if we plot $\left\langle \overline{V_\times} \right\rangle$ as a 
function of $\left< k \right>$, curves corresponding to different parameter combinations $(n,\sigma,\ell)$ 
will fall on top of the universal curve given by Eq.~(\ref{Vxofk}). Indeed, in Fig.~\ref{Fig05} we present
the curves of $\left< V_\times \right>$ (divided by $n$) of Fig.~\ref{Fig02}(a-c) but now as a function of 
$\left< k \right>$ and observe, as expected, that all curves $\left\langle \overline{V_\times} \right\rangle$ 
vs.~$\left< k \right>$ fall one on top of the other (except for small differences in the interval 
$1<\left< k \right><10$, see the insets; these differences are observed when $\sigma<1$).
In all panels we also plot Eq.~(\ref{Vxofk}) as dashed lines and observe a very good 
correspondence with the numerical data, which is quite remarkable since Eq.~(\ref{Vxofk})
was expected to work only in the limit of $L\ll 1$.

The scaling of $\left< V_\times \right>/n$ of non-uniform RGGs with the average degree, see 
Fig.~\ref{Fig05}, agrees with the scaling of several (normalized) topological indices with 
$\left< k \right>$ as reported in Refs.~\cite{AHMS20,AMRS20} for Erd\"os-Renyi graphs and 
RGGs in the unit square. However, here we are providing an explicit expression for the scaling, 
see Eq.~(\ref{Vxofk}). Moreover, we expect other topological indices on non-uniform RGGs
to be also scale invariant with $\left< k \right>$; see the next Subsection, where it is shown that the
Randi\'c connectivity index also scales with $\left< k \right>$.

\subsection{Randi\'c connectivity index}
\label{Sub:R}

As already mentioned in the previous Subsection, the scaling of $\left< V_\times \right>/n$ of non-uniform 
RGGs with the average degree, see Fig.~\ref{Fig05}, make us expect that other topological indices 
on non-uniform RGGs to be also scale invariant with $\left< k \right>$.
Thus, in the following we verify this expectation by the use of the Randi\'c connectivity index $R$.

The \emph{Randi\'c connectivity index} was defined in~\cite{R} as
\begin{equation}
\label{R}
R = \sum_{uv} \frac1{\sqrt{d_u d_v}}\, ,
\end{equation}
where $uv$ denotes the edge of the graph connecting the vertices $u$ and $v$ and $d_u$ is the 
degree of the vertex $u$. 
In addition to the multiple applications of the Randi\'c index in physical chemistry, being one of the 
most popular topological indices (see, e.g.,~\cite{GF,LG,LS} and the references therein),
this index has found several applications in other research areas and topics, such as information 
theory~\cite{GFK18}, network similarity~\cite{NJ03}, protein alignment~\cite{R15}, network 
heterogeneity~\cite{E10}, and network robustness~\cite{MMR17}. However, its use in the study
of random graphs has been scarce. For recent exceptions see Refs.~\cite{MMRS20,AHMS20,AMRS20}, 
where the average Randi\'c index has been used to probe the percolation transition in Erd\"os-R\'enyi 
graphs and RGGs.

\begin{figure*}
\begin{center} 
\includegraphics[width=0.8\textwidth]{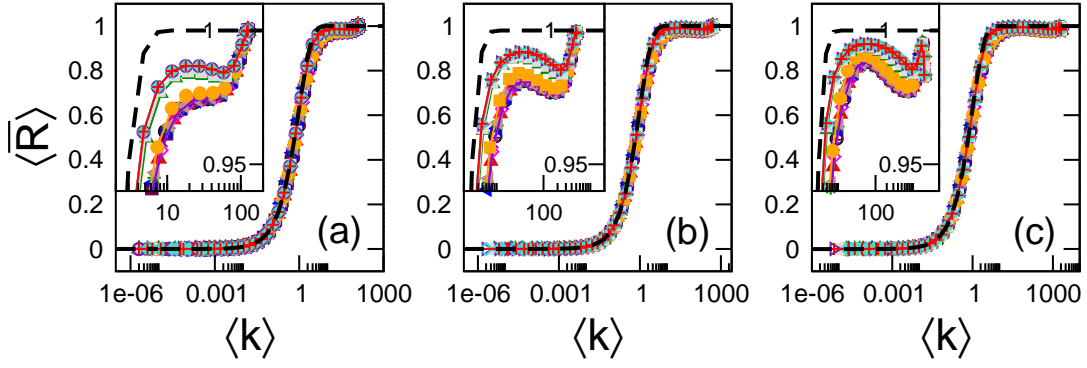}
\caption{
Average Randi\'c index $\left\langle R \right\rangle$ normalized to $n/2$ for (a) $n=125$, (b) $n=500$ and (c) 
$n=2000$ as a function of the average degree $\left\langle k \right\rangle$ of non-uniform RGGs.
Each panel displays 14 curves corresponding to different degrees of non-uniformity 
$\sigma$: $\{0.001, 0.002, 0.004, 0.01, 0.02, 0.04, 0.1, 0.2, 0.4, 0.8, 1, 10, 100, \infty \}$.
Dashed line in all panels is Eq.~(\ref{Rofk}).
Each data value was computed by averaging over $10^7/n$ random graphs.
}
\label{Fig06}
\end{center}
\end{figure*}

In Fig.~\ref{Fig06} we present the normalized average Randi\'c index $\left\langle \overline{R} \right\rangle$ as a 
function of the average degree $\left\langle k \right\rangle$ of non-uniform RGGs of size $n$. As for $\left\langle V_\times \right\rangle$, 
we normalize $\left\langle R \right\rangle$ to the maximum value it can take:
$\left\langle \overline{R} \right\rangle \equiv \left\langle R \right\rangle/\left\langle R \right\rangle_{\tbox{GOE}}$, with $\left\langle R \right\rangle_{\tbox{GOE}}=n/2$. 
As anticipated, we observe that
$\left\langle \overline{R} \right\rangle$ is properly scaled with $\left\langle k \right\rangle$, except for the region of large $\left\langle k \right\rangle$
(see the insets) where we observe two sets of curves falling one on top of the other: 
one set corresponding to $\sigma<1$ and the other to $\sigma\ge 1$. This effect is equivalent to that
observed for $\left\langle \overline{V_\times} \right\rangle$ (see the insets of Fig.~\ref{Fig05}).

\begin{figure*}
\begin{center} 
\includegraphics[width=0.7\textwidth]{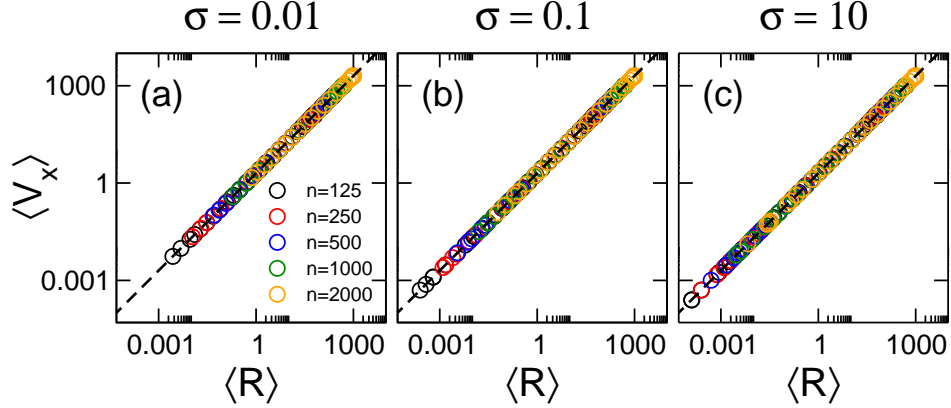}
\caption{
Average number of non-isolated vertices $\left\langle V_\times \right\rangle$ as a function of the average Randi\'c 
index $\left\langle R \right\rangle$ of non-uniform RGGs of several sizes $n$ characterized by (a) $\sigma=0.01$, 
(b) $\sigma=0.1$ and (c) $\sigma=10$. The dashed line on top of the data is 
$\left\langle V_\times \right\rangle=2\left\langle R \right\rangle$.
Each data value was computed by averaging over $10^7/n$ random graphs.
}
\label{Fig07}
\end{center}
\end{figure*}

At this point it is relevant to recall that in Ref.~\cite{AMRS20} it was shown that $\left\langle V_\times \right\rangle$
and $\left\langle R \right\rangle$ on RGGs are highly correlated, which also occurs for non-uniform RGGs; as can 
be clearly seen in Figs.~\ref{Fig07}(a,b) where we plot $\left\langle V_\times \right\rangle$ vs.~$\left\langle R \right\rangle$ for
non-uniform RGGs with $\sigma<1$. Moreover, Fig.~\ref{Fig07} also suggest that
\begin{equation}
\left\langle V_\times \right\rangle \approx 2 \left\langle R \right\rangle ,
\label{VvsR} 
\end{equation}
see the dashed lines on top of the data in Fig.~\ref{Fig07}.
Thus, Eq.~(\ref{VvsR}) in addition to Eq.~(\ref{Vxofk}), allows us to propose
\begin{equation}
\left< R \right> \approx (n/2)\left[1-\exp(-\left< k \right>)\right] \, ,
\label{Rofk}
\end{equation}
which in fact coincides relatively well with the numerical data reported in Fig.~\ref{Fig06} (see
the dashed lines); except for the region of large $\left\langle k \right\rangle$ where significant differences between
Eq.~(\ref{Rofk}) and the numerical data are evident (see the insets).

\begin{figure*}
\begin{center} 
\includegraphics[width=0.8\textwidth]{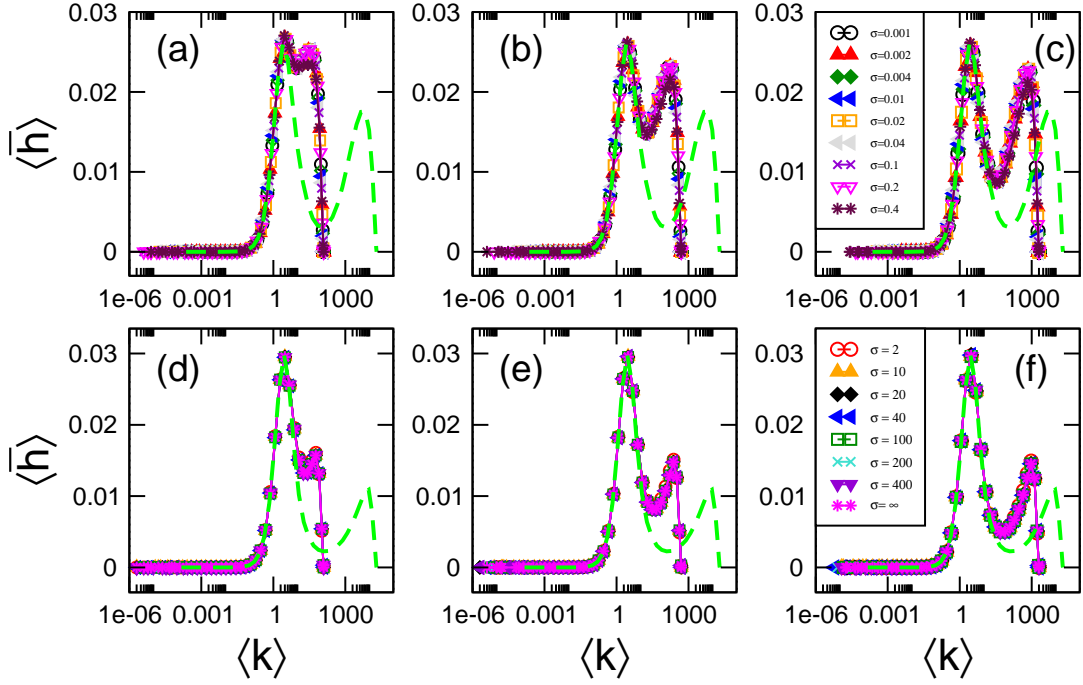}
\caption{
Average heterogeneity index $\left\langle h \right\rangle$ normalized to $n$ for (a,d) $n=125$, (b,e) $n=500$ 
and (c,f) $n=2000$ as a function of the average degree $\left\langle k \right\rangle$ of non-uniform RGGs
with (a-c) $\sigma<1$ and (d-f) $\sigma\ge 1$. The green dashed line in (d-f) [(g-i)] corresponds 
to $n=20,000$ and $\sigma=0.4$ [$n=20,000$ and $\sigma=10$].
Each data value was computed by averaging over $10^7/n$ random graphs.)
}
\label{Fig08}
\end{center}
\end{figure*}

It is fair to admit that the log-log scale we used to present the data in Fig.~\ref{Fig07} makes the  
approximation of Eq.~(\ref{VvsR}) to look very accurate, but it is not.
Then, to quantify the accuracy of Eq.~(\ref{VvsR}) we will make use of the heterogeneity index~\cite{E10}
\begin{equation}
\label{h}
h = \sum_{uv} \left( \frac1{\sqrt{d_u}} - \frac1{\sqrt{d_v}} \right)^2 \, ,
\end{equation}
which can be written in terms of $V_\times$ and $R$ as:
\begin{equation}
\label{h2}
h = \sum_{uv} \left( \frac1{d_u} + \frac1{d_v} \right) - 2 \sum_{uv} \left( \frac1{\sqrt{d_ud_v}} \right) = 
V_\times - 2R \, .
\end{equation}
Note that Eq.~(\ref{VvsR}) implies $\left< h \right> \approx 0$ for any combination of parameters 
$(n,\sigma,\ell)$. Nevertheless, as clearly shown in Fig.~\ref{Fig08} (where we plot $\left< h \right>/n$ 
vs.~$\left\langle k \right\rangle$ for non-uniform RGGs of different sizes and non-uniformity strengths $\sigma$), 
the curves $\left\langle \overline{h} \right\rangle$ vs.~$\left\langle k \right\rangle$ develop a two-peak structure for 
$\left\langle k \right\rangle \stackrel{>}{\sim} 1$ with maxima closer to 0.03. Also note that the two-peak structure
changes with $n$, differently for $\sigma<1$ and $\sigma\ge 1$, making $\left\langle \overline{h} \right\rangle$
non-scalable.
 
\subsection{Ratio of consecutive eigenvalue spacings and Shannon entropy}
\label{Sub:spectral}

Once we have concluded that the average degree is the scaling parameter of 
$\left\langle \overline{V_\times} \right\rangle$, our first conjecture is that $\left< k \right>$ may also be the scaling 
parameter of $\left< \overline{r} \right>$ and $\left< \overline{S} \right>$.

First we normalize $\left< r \right>$ and $\left< S \right>$, so that we can compare them for different graph 
sizes $n$. We naturally choose $\left\langle \overline{S} \right\rangle=\left< S \right>/\left< S \right>_{\tbox{GOE}}$ with
$\left< S \right>_{\tbox{GOE}}\approx \ln (n/2.07)$~\cite{MK98}, however the
small-size effects observed for $\left< r \right>$ and the fact that $\left< r \right>\to\mbox{const.}\ne 0$
when $\ell\to 0$ make us conveniently define $\left\langle \overline{r} \right\rangle$ as
$$
\left\langle \overline{r} \right\rangle = \frac{\left< r \right>-\left< r(\ell=0) \right>}{\left< r(\ell=2) \right>-\left< r(\ell=0) \right>} \ ;
$$
where $\left< r(\ell=0) \right>$ and $\left< r(\ell=2) \right>$, which do not depend on $\sigma$, are 
numerically computed for a given $n$. Evidently,
$\left< r(\ell=0) \right>\to\left< r \right>_{\tbox{PE}}\approx 0.3863$~\cite{ABG13} and
$\left< r(\ell=2) \right>\to\left< r \right>_{\tbox{GOE}}\approx 0.5359$~\cite{ABG13} for large enough $n$.

\begin{figure*}[t!]
\begin{center} 
\includegraphics[width=0.75\textwidth]{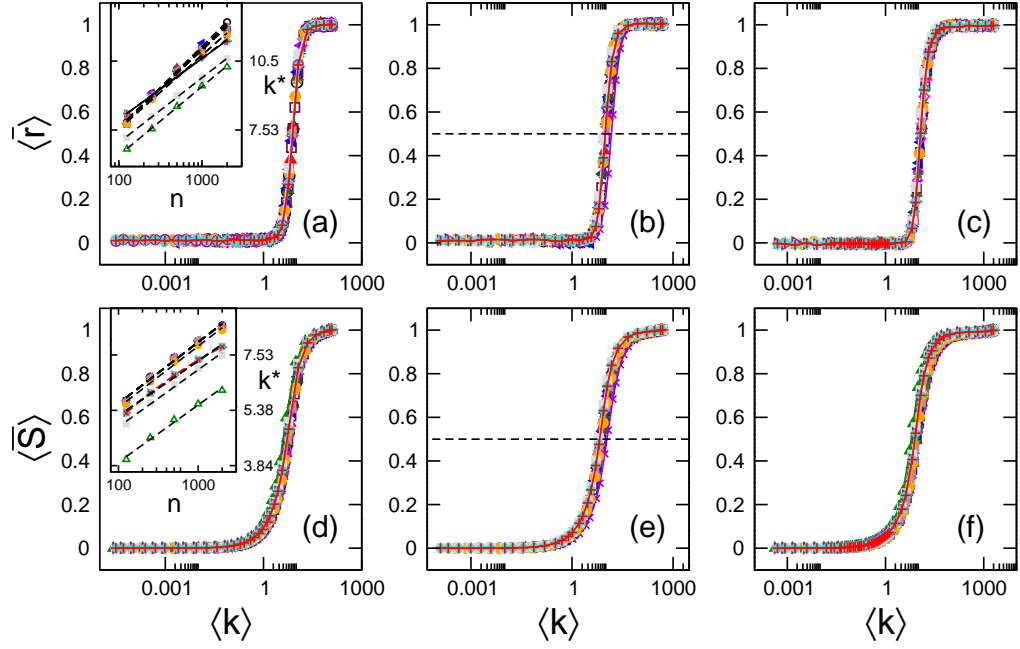}
\caption{
Normalized (a-c) average ratio of consecutive eigenvalue spacings $\left< \overline{r} \right>$ and
(d-f) average Shannon entropy $\left< \overline{S} \right>$ as a function of the 
average degree $\left< k \right>$ of non-uniform RGGs of size (a,d) $n=125$, (b,d) $n=500$, and 
(c,f) $n=2000$. Same data as in Fig.~\ref{Fig02}(d-i). The insets in panels (a) and (d) show $k^*$ 
vs.~$n$ as extracted from the intersection of the curves $\left\langle \overline{r} \right\rangle$ vs.~$\left\langle k \right\rangle$ 
and $\left\langle \overline{S} \right\rangle$ vs.~$\left\langle k \right\rangle$ with the straight lines $\left\langle \overline{r} \right\rangle = 0.5$ 
and $\left\langle \overline{S} \right\rangle = 0.5$, respectively. Dashed lines are fittings
of the data with Eq.~(\ref{scalingk}). The exponents $\gamma$ obtained from the fittings are 
reported in Table~\ref{Table0}. Horizontal dashed lines in panels (b,e) mark $\left\langle \overline{X} \right\rangle = 0.5$.
}
\label{Fig09}
\end{center}
\end{figure*}

Then, in Fig.~\ref{Fig09} we present $\left< \overline{r} \right>$ and $\left< \overline{S} \right>$ as a function 
of $\left\langle k \right\rangle$; note that the data shown in Fig.~\ref{Fig09} is the same as that of Fig.~\ref{Fig02}(d-i).
Even though, from this figure we can clearly see that the curves $\left\langle \overline{X} \right\rangle$ vs.~$\left\langle k \right\rangle$
fall one on top of the other in each of the figure panels, there is still a small but measurable dependence
of these curves on $n$. That is, while they keep their functional form, they suffer a displacement on the 
$\left\langle k \right\rangle$-axis by increasing $n$. Therefore, we conclude that $\left\langle k \right\rangle$ does not scale the spectral
nor eigenvector properties of our random graph model. Thus, in order to search for the proper scaling 
parameter $\xi$ we first establish a quantity to characterize the 
position of the curves $\left\langle \overline{X} \right\rangle$ on the $\left\langle k \right\rangle$-axis.
Since all curves $\left\langle \overline{X} \right\rangle$ vs.~$\left\langle k \right\rangle$ transit from zero (PE regime) to one 
(GOE regime) when $\left\langle k \right\rangle$ increases from small to large values, we choose the value of 
$\left\langle k \right\rangle$ for which $\left\langle \overline{X} \right\rangle \approx 0.5$; see the horizontal dashed lines in 
Fig.~\ref{Fig09}(b,e). We label the value of $\left\langle k \right\rangle$ at half of 
the PE to GOE transition as $k^*$.

\begin{table*}
\begin{center}
\scalebox{0.95}{
\begin{tabular}{l|l|l|l|l|l|l|l|l|l|l|l|l|l|l|l|l|l|l|l|}
\cline{2-14}
 &$\sigma=0.001$ & $\sigma=0.002$ & $\sigma=0.004$ & $\sigma=0.01$ & $\sigma=0.02$ & $\sigma=0.04$ & $\sigma=0.1$ & $\sigma=0.2$ & $\sigma=0.4$ & $\sigma=0.8$& $\sigma=1$ & $\sigma=\infty$ & $\left\langle \gamma \right\rangle$   \\ \hline
\multicolumn{1}{|l|}{$k^*(r)$ vs.~$n$}
&0.1739 &0.1656 &0.1659 &0.1691 &0.1622 &0.1672 &0.1693 &0.1565 &0.1461 &0.1376 &0.1307 & 0.1297 & 0.1561\\ \hline
\multicolumn{1}{|l|}{$k^*(S)$ vs.~$n$}
&0.1626 & 0.1619 & 0.1614 & 0.1614 & 0.1624 & 0.1613 &
0.1621 & 0.1619&0.1509 & 0.1508 & 0.1443 &0.1455 &0.1572
\\ \hline
\end{tabular}}
\caption{Values of the exponent $\gamma$ obtained from the fittings of the curves $k^*(X)$ vs.~$n$ 
of the insets in Fig.~\ref{Fig09} with Eq.~(\ref{scalingk}). The average value of $\gamma$ is reported 
in the right-most column.} 
\label{Table0}
\end{center}
\end{table*}

In the insets of Figs.~\ref{Fig09}(a) and~\ref{Fig09}(d) we report $k^*$ vs.~$n$ as extracted from 
the intersection of the curves $\left\langle \overline{r} \right\rangle$ vs.~$\left\langle k \right\rangle$ and $\left\langle \overline{S} \right\rangle$ vs.~$\left\langle k \right\rangle$
with the straight lines $\left\langle \overline{r} \right\rangle = 0.5$ and $\left\langle \overline{S} \right\rangle = 0.5$, 
respectively. Indeed, the linear trend of the data sets (in log-log scale) $k^*$ vs.~$n$ suggests the power-law 
behavior
\begin{equation}
\label{scalingk}
k^* = \mathcal{C}n^\gamma .
\end{equation}
As shown in the insets of Figs.~\ref{Fig09}(a) and~\ref{Fig09}(d), Eq.~(\ref{scalingk}) provides excellent 
fittings to the data; see the dashed lines.
From the fitted exponents, reported in Table~\ref{Table0}, we can conclude that
$\gamma \approx 0.16$ for both the ratio of consecutive eigenvalue spacings and Shannon entropy, 
for all values of $\sigma$.

\begin{figure*}[t!]
\begin{center} 
\includegraphics[width=0.6\textwidth]{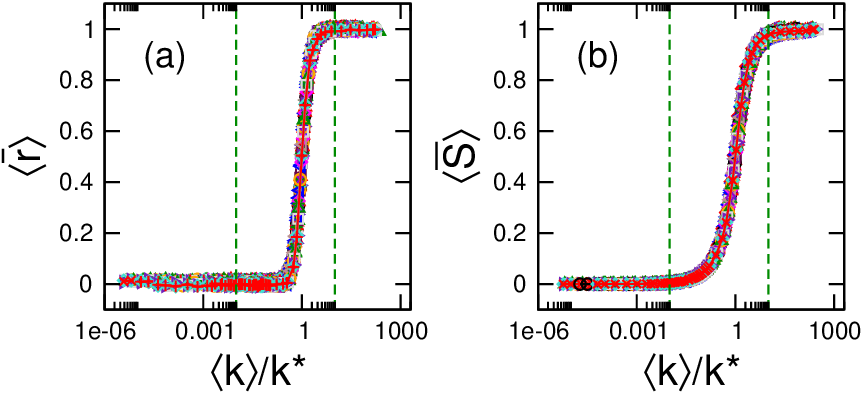}
\caption{
(a) $\left< \overline{r} \right>$ and (b) $\left< \overline{S} \right>$ as a function of the scaling
parameter $\xi=\left< k \right>/k^*$. Same data as in Fig.~\ref{Fig09}.
Vertical dashed lines at $\xi=0.01$ and 10 mark, approximately, the onset of 
delocalization and the onset of the GOE regime, respectively.
}
\label{Fig10}
\end{center}
\end{figure*}

Finally, we define the scaling parameter as the ratio between $\left< k \right>$ and $k^*$, so we get  
\begin{equation}
\label{xi}
\xi \equiv \frac{\left< k \right>}{k^*} \propto \frac{\left< k \right>}{n^\gamma} = n^{-\gamma}\left< k \right> .
\end{equation} 
Therefore, by plotting again the curves of $\left\langle \overline{X} \right\rangle$ now as a function of $\xi$
we observe that curves for different graph sizes $n$ and non-uniformity strengths $\sigma$ collapse 
on top of {\it universal curves}; see Fig.~\ref{Fig10}. 
Also note that each measure $X$ is characterized by a slightly different universal curve.
In particular we observe that the PE-to-GOE transition is sharper for $\left\langle \overline{r} \right\rangle$, as compared 
to $\left\langle \overline{S} \right\rangle$.

\subsection{Straightforward scaling of $V_\times$, $r$ and $S$}
\label{Sub:straight}

Above we performed the scaling analysis of $V_\times$, $r$ and $S$ separately, first for 
$V_\times$ in Sect.~4.1 and later for $r$ and $S$ in Sect.~4.2.
In both cases we took advantage of the previous knowledge of a heuristic expression for $\left\langle k \right\rangle$, 
see Eqs.~(\ref{k},\ref{f(L)}).
However, in other works we have successfully performed scaling studies of both topological and spectral
properties of random graph models without any previous insight about the functional form of the scaling 
parameter, see e.g.~\cite{MAMPS19,MAM15,AMGM18,MFMR17}. 
Thus, in this Subsection we perform a straightforward scaling analysis of the three measures $V_\times$, $r$ 
and $S$ and show that we obtain equivalent results as those reported in Sects.~4.1 and~4.2.

\begin{figure*}
\begin{center} 
\includegraphics[width=0.85\textwidth]{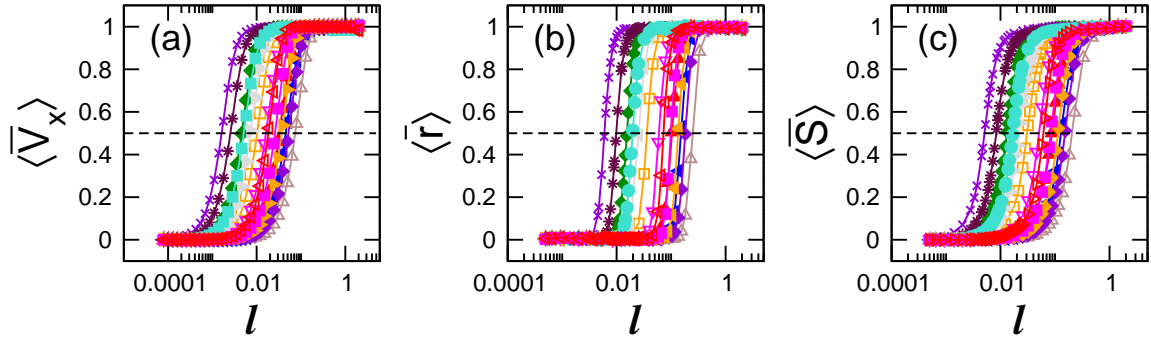}
\caption{
Normalized (a) number of non-isolated vertices $\left\langle \overline{V_\times} \right\rangle$,
(b) average ratio of consecutive eigenvalue spacings $\left< \overline{r} \right>$, and
(c) average Shannon entropy $\left< \overline{S} \right>$
as a function of the connection radius $\ell$ of non-uniform RGGs.
Each panel displays 14 curves corresponding to the following combinations of  
$(\sigma,n)$: $\{(0.002,1000),(0.01,2000),(0.004,250),(0.04,2000),$ $(0.02,500),(0.04,500),
(0.2,1000),(0.8,2000),(0.8,1000),(0.1,125),(0.8,500),(0.4,250),$ $(0.8,250), (0.8,125) \}$, form left to right.
The dashed lines at $\left\langle \overline{X} \right\rangle = 0.5$ are used to extract $\ell^*$, see the text.
Each data value was computed by averaging over $10^7/n$ random graphs.
}
\label{Fig11}
\end{center}
\end{figure*}

Taking as a starting point the observations (i-v) made in Sect.~2.3 from Fig.~\ref{Fig02}, in Fig.~\ref{Fig11} 
we present again the measures $\left\langle X \right\rangle$ but now they are conveniently normalized as in Figs.~\ref{Fig05} 
and~\ref{Fig09}.
Note that some of the curves presented in Fig.~\ref{Fig11} were already reported in Fig.~\ref{Fig02},
however we are including curves corresponding to additional parameter combinations.
From this figure we can clearly see that when changing $n$ and $\sigma$ the curves 
$\left\langle \overline{X} \right\rangle$ keep their functional form but they suffer a displacement on the 
$\ell$-axis. Therefore, in order to search for the scaling parameter $\chi\equiv \chi(n,\sigma,\ell)$ 
we first establish a quantity to characterize the 
position of the curves $\left\langle \overline{X} \right\rangle$ on the $\ell$-axis.
Since all curves $\left\langle \overline{X} \right\rangle$ vs.~$\ell$ transit from zero (PE regime) to one 
(GOE regime) when $\ell$ increases from zero to two, we choose the value of 
$\ell$ for which $\left\langle \overline{X} \right\rangle \approx 0.5$; see the horizontal dashed lines in 
Fig.~\ref{Fig11}. We label the value of $\ell$ at half of 
the PE to GOE transition as $\ell^*$; so we call $\ell^*$ the PE-to-GOE transition point.
\begin{figure*}
\begin{center} 
\includegraphics[width=0.8\textwidth]{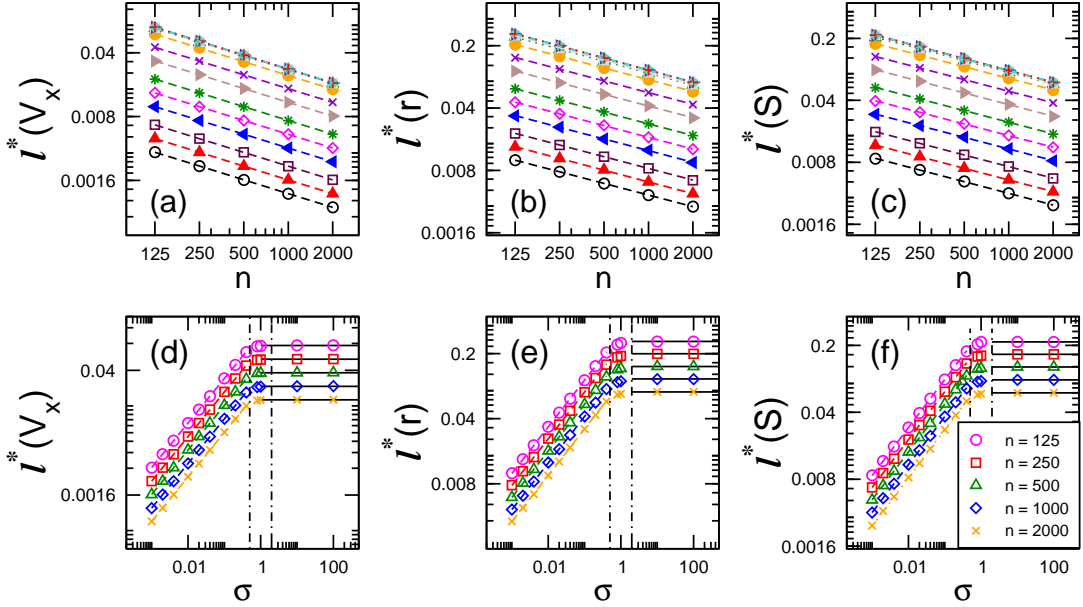}
\caption{
PE-to-GOE transition point $\ell^*$ as a function of (a-c) the graph size $n$ and (d-f) the 
non-uniformity $\sigma$ fom (a,d) the number of non-isolated vertices, (b,e) the ratio of 
consecutive eigenvalue spacings, and (c,f) the Shannon entropy. 
Several values of $\sigma$ [$n$] are reported in the upper [lower] panels. Each of the 
upper panels (a-c) displays 14 data sets corresponding to different degrees of non-uniformity 
$\sigma$: $\{0.001, 0.002, 0.004, 0.01, 0.02, 0.04, 0.1, 0.2, 0.4, 0.8, 1, 10, 100, \infty \}$, 
form bottom to top.
The dashed lines are fittings to the data with Eq.~(\ref{scaling}); the fitted exponents
$\gamma_\ell$ and $\delta$ are reported in Tables~\ref{Table1} and~\ref{Table2}, respectively.
Dash-dotted vertical lines in the lower panels indicate a transition region around $\sigma_c\approx1$.
Full lines in the lower panels on top of the data for $\sigma>1$ correspond to the value of 
$\ell^*$ at $\sigma\to\infty$.
}
\label{Fig12}
\end{center}
\end{figure*}

Given that  $\ell^*$ depends on both $n$ and $\sigma$, in Fig.~\ref{Fig12} we report $\ell^*$ versus 
$n$ for fixed values of $\sigma$ (upper panels) and $\ell^*$ versus $\sigma$ for fixed values of $n$ 
(lower panels). It is interesting to notice that while $\ell^*$ decreases as a function of $n$ (see the 
upper panels of Fig.~\ref{Fig12}) for all the values of $\sigma$ reported here, the curves $\ell^*$ 
vs.~$\sigma$ show two different behaviors (see the lower panels of Fig.~\ref{Fig12}):
for $\sigma<1$, $\ell^*$ grows with $\sigma$ but when $\sigma>1$, $\ell^*\approx\mbox{const.}$,
with a transition region around $\sigma_c\approx 1$.
Therefore, we define two scaling regimes: {\it the clustering regime} when $\sigma<\sigma_c$ and 
{\it the uniform regime} when $\sigma>\sigma_c$. 
Note that since $\sigma$ is given in units of the disc radius, $\sigma_c$ can be indeed interpreted 
as the disc radius. These two regimes are exemplified graphically in 
Fig.~\ref{Fig01}. In fact, once $\sigma>\sigma_c$ our random graph model already reproduces
the random geometric graph model on the disc. Moreover, the full horizontal lines on top of the data 
of Fig.~\ref{Fig12}(d-f) for $\sigma>\sigma_c$ corresponds to the value of $\ell^*$ at 
$\sigma\to\infty$; that is, once $\sigma>\sigma_c$, the properties of our random graph model do 
not change anymore by further increasing $\sigma$; as already noticed in Fig.~\ref{Fig02}.

Indeed, the linear trend of the data sets (in log-log scale) $\ell^*$ vs.~$n$ and $\ell^*$ vs.~$\sigma$ 
suggests the power-law behaviors
\begin{equation}
\label{scaling}
\ell^* = \mathcal{C}n^{-\gamma_\ell}\sigma^\delta .
\end{equation}
As shown in Fig.~\ref{Fig12}, Eq.~(\ref{scaling}) provides excellent fittings to the data; see the dashed 
lines.
From the fitted exponents, reported in Tables~\ref{Table1} and~\ref{Table2}, we can safely state that:
$\gamma_\ell\approx 1/2$ for the number of non-isolated vertices, while $\gamma_\ell\approx 0.43$ for both
the ratio of consecutive eigenvalue spacings and Shannon entropy, for all values of $\sigma$. 
Also, $\delta\approx 1/2$ for $\sigma<\sigma_c$ while $\delta\approx 0$ for $\sigma>\sigma_c$, 
for all the three measures ($V_\times$, $r$ and $S$).

\begin{table*}
\begin{center}
\scalebox{0.9}{
\begin{tabular}{l|l|l|l|l|l|l|l|l|l|l|l|l|l|l|l|}
\cline{2-11}
 &$\sigma=0.001$ & $\sigma=0.002$ & $\sigma=0.004$ & $\sigma=0.01$ & $\sigma=0.02$ & $\sigma=0.04$ & $\sigma=0.1$ & $\sigma=0.2$ & $\sigma=0.4$ & $\left\langle \gamma_\ell \right\rangle$   \\ \hline
\multicolumn{1}{|l|}{$\ell^*(V_x)$ vs.~$n$}  &0.5010& 0.5015 &0.5007 &0.5008& 0.5013 &0.5008 &0.5012 &0.5012 &0.5007 &0.5010\\ \hline
\multicolumn{1}{|l|}{$\ell^*(r)$ vs.~$n$}
&0.4316 &0.4355 &0.4353 &0.4333 &0.4369 &0.4350 &0.4343 &0.4364 &0.4375 &0.4350\\ \hline
\multicolumn{1}{|l|}{$\ell^*(S)$ vs.~$n$}  &0.4364& 0.4345 &0.4348 &0.4344 &0.4340 &0.4346 &0.4342 &0.4336 &0.4331 &0.4344\\ \hline
\end{tabular}}
\caption{Values of the exponent $\gamma_\ell$ obtained from the fittings of the curves $\ell^*(X)$ vs.~$n$ 
of Fig.~\ref{Fig12}(a-c) with Eq.~(\ref{scaling}). The average value of $\gamma_\ell$ is reported in the right-most column.} 
\label{Table1}
\end{center}
\end{table*}

\begin{table}
\begin{center}
\scalebox{0.9}{
\begin{tabular}{l|l|l|l|l|l|l|}
\cline{2-7}
& $n=125$ & $n=250$ & $n=500$ & $n=1000$ & $n=2000$ & $\left\langle \delta \right\rangle$  \\ \hline
\multicolumn{1}{|l|}{$\ell^*(V_{x})$ vs.~$\sigma$} & 0.4987  & 0.4989     & 0.4983    & 0.4991 & 0.4987 & 0.4987\\ \hline
\multicolumn{1}{|l|}{$\ell^*(r)$ vs.~$\sigma$} & 0.5013     & 0.5016   & 0.5007     & 0.4994 & 0.4993 & 0.5005 \\ \hline
\multicolumn{1}{|l|}{$\ell^*(S)$ vs.~$\sigma$} & 0.5018  & 0.5007     & 0.4996     & 0.5012 & 0.5025& 0.5011 \\ \hline
\end{tabular}}
\caption{Values of the exponent $\delta$ obtained from the fittings of the curves $\ell^*(X)$ vs.~$\sigma$
(for $\sigma<1$) of Fig.~\ref{Fig12}(d-f) with Eq.~(\ref{scaling}). The average value of $\delta$ is reported 
in the right-most column.} 
\label{Table2}
\end{center}
\end{table}

\begin{figure*}
\begin{center} 
\includegraphics[width=0.85\textwidth]{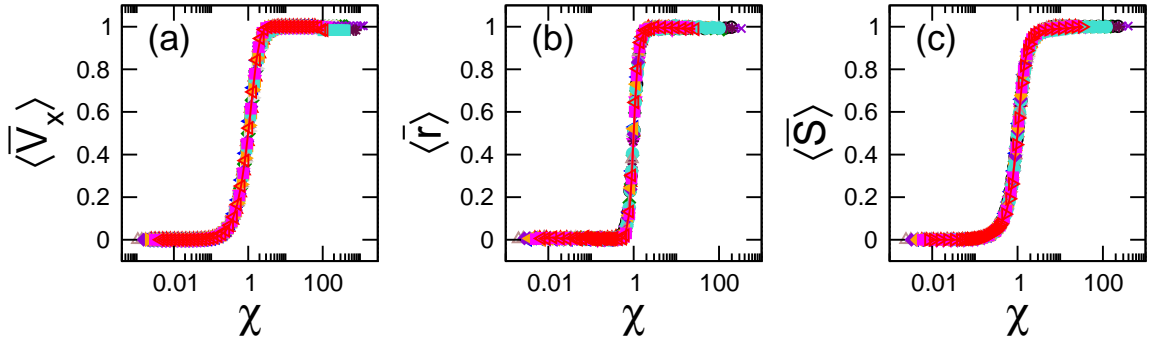}
\caption{
Normalized (a) number of non-isolated vertices $\left\langle \overline{V_\times} \right\rangle$,
(b) average ratio of consecutive eigenvalue spacings $\left< \overline{r} \right>$, and
(c) average Shannon entropy $\left< \overline{S} \right>$
as a function of the scaling parameter $\chi$, see Eq.~(\ref{chi}), of non-uniform RGGs.
Same curves as in Fig.~\ref{Fig05}.
}
\label{Fig13}
\end{center}
\end{figure*}

Finally, we define the scaling parameter $\chi$ as the ratio between $\ell$ and $\ell^*$, so we get  
\begin{equation}
\label{chi}
\chi \equiv \frac{\ell}{\ell^*} \propto \frac{\ell}{n^{-\gamma_\ell}\sigma^\delta} = n^{\gamma_\ell}\sigma^{-\delta}\ell \ .
\end{equation} 
Therefore, by plotting again the curves of $\left\langle \overline{X} \right\rangle$ now as a function of $\chi$
we observe that curves for different graph sizes $n$ and non-uniformity strengths $\sigma$ collapse 
on top of {\it universal curves}; see Fig.~\ref{Fig13}. 

It is fair to mention that the scaling we found for $\left\langle \overline{S} \right\rangle$ when $\sigma>\sigma_c$
is very close to that reported in~\cite{AMGM18} for RGGs in the unit square, as expected.
There, $\chi\equiv\chi(n,\ell) \propto n^{\gamma_\ell}\ell$ with $\gamma_\ell\approx 0.425$. 

At first sight, it seems that in this Subsection we got different scaling parameters than in the previous one:
On the one hand, recall that in Subsection~\ref{Sub:spectral} we found 
$\xi=\left\langle k \right\rangle\propto n\sigma^{-1}\ell^2$ for $\left\langle \overline{V_\times} \right\rangle$, 
while we got $\xi=n^{-0.16}\left\langle k \right\rangle\propto n^{0.84}\sigma^{-1}\ell^2$ for $\left\langle \overline{r} \right\rangle$ and 
$\left\langle \overline{S} \right\rangle$ (here we are using 
$\left\langle k \right\rangle \propto n\sigma^{-1}\ell^2$ when $L\ll 1$ and $\sigma<\sigma_c$).
On the other hand in this Subsection we have obtained 
$\chi \propto n^{1/2}\sigma^{-1/2}\ell$ for $\left\langle \overline{V_\times} \right\rangle$ and 
$\chi \propto n^{0.43}\sigma^{-1/2}\ell$ for $\left\langle \overline{r} \right\rangle$ and 
$\left\langle \overline{S} \right\rangle$.
This apparent mismatch can be understood by noticing that not only $\xi$ but any function of it should 
scale the normalized measures $\left\langle \overline{X} \right\rangle$; thus, since $\chi\propto\xi^{1/2}$, for the three 
measures, our results are consistent.

\section{Discussion and conclusions}

We performed a detailed scaling study of random geometric graphs (RRGs) in the unit disc 
characterized by a non-uniform density of vertices. This random graph model may serve as a reference
model of complex systems embedded in the plane whose components are not uniformly allocated. 
Our random graph model depends on three parameters: the number of vertices $n$, the degree of 
non-uniformity $\sigma\in(0,\infty)$ and the connection radius $\ell\in[0,2]$.
This model produces a cluster around the disc center for $\sigma<\sigma_c$ and reproduces the uniform 
RGG model in the disc when $\sigma\ge \sigma_c$ (see Fig.~\ref{Fig01} and Subsection~\ref{Sub:straight}) with 
$\sigma_c\approx 1$.

By the use of the average degree $\left\langle k \right\rangle$, the number non-isolated vertices $V_\times$, the ratio 
of consecutive eigenvalue spacings $r$ and the Shannon entropy $S$ of eigenvectors we probe topological 
as well as spectral properties of our random graph model.
First we propose a heuristic expression able to properly describe $\left\langle k(n,\sigma,\ell) \right\rangle$; see 
Eqs.~(\ref{k},\ref{f(L)}). 
Then, we looked for the scaling properties of the properly normalized average measure 
$\left\langle \overline{X} \right\rangle$ (where $X$ stands for $V_\times$, $r$ and $S$).
As a result of the scaling analysis, we were able to define the scaling parameter, that we label $\xi$, 
such that the curves $\left\langle \overline{X} \right\rangle$ vs.~$\xi$ are invariant curves.
Particularly, in the two graph regimes separated by the critical non-uniformity $\sigma_c$, we found that 
$\xi=\left\langle k \right\rangle$ for $\left\langle \overline{V_\times} \right\rangle$ while $\xi=n^{-\gamma}\left\langle k \right\rangle$, with 
$\gamma\approx 0.16$, for $\left\langle \overline{r} \right\rangle$ and $\left\langle \overline{S} \right\rangle$. 
In addition, we found that $\left\langle \overline{V_\times} \right\rangle=\left\langle V_\times \right\rangle/n$ is related to $\left\langle k \right\rangle$ 
as $\left\langle \overline{V_\times} \right\rangle\approx 1-\exp(-\left\langle k \right\rangle)$, see Eq.~(\ref{Vxofk})
and Fig.~\ref{Fig05}.

We stress that the scalings shown in Figs.~\ref{Fig05} and~\ref{Fig10} have two important consequences 
in the characterization of our non-uniform random graph model.
First, they allow us to define regimes: 
The PE [GOE] regime can be defined for $\xi<0.01$ [$\xi>10$], while 
$0.01<\xi<10$ defines the PE-to-GOE transition regime.
Here, the PE regime is characterized by mostly disconnected vertices and localized eigenvectors
while the GOE regime corresponds to almost complete graphs and delocalized eigenvectors.
Thus, $\xi = 0.01$ and $\xi=10$ (see the vertical dashed lines in Figs.~\ref{Fig05} and~\ref{Fig10}) mark, 
approximately, the percolation transition, or the onset of eigenvector delocalization, and the onset of the 
GOE limit, respectively.
Second, it allow us to make predictions: 
Given a combination of parameters $(n,\sigma,\ell)$,
if  $\xi<0.01$ we know that $\left\langle V_\times \right\rangle\approx 0$, $\left\langle r \right\rangle\approx 0.3863$ and  
$\left\langle S \right\rangle\approx 0$; while if $\xi>10$ we expect $\left\langle V_\times \right\rangle\approx n$, 
$\left\langle r \right\rangle\approx 0.5359$ and  $\left\langle S \right\rangle\approx \ln(n/2.07)$.

We also want to note that the number of non-isolated vertices, as well as the Randi\'c connectivity index 
(see Subsection~\ref{Sub:R}), have provided us with equivalent information than 
standard RMT measures, that is, we were able to clearly identify both the PE and the GOE regimes,
as well as the PE-to-GOE transition regime, by means of the {\it universal} curves of 
$\left\langle \overline{V_\times} \right\rangle$ and $\left\langle \overline{R} \right\rangle$ vs.~$\xi$; thus 
we give further evidence of the usefulness of topological indices
in the statistical characterization of random graphs. 

\begin{figure*}
\begin{center} 
\includegraphics[width=0.75\textwidth]{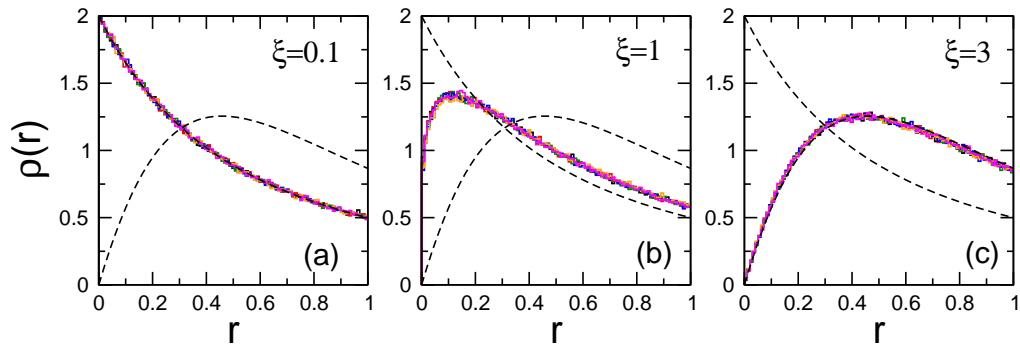}
\caption{
Histograms of the probability distribution function of the ratio of consecutive eigenvalue spacings 
$\rho(r)$ for three values of the scaling parameter $\xi$. 
Each panel displays 6 histograms corresponding to the following combinations of  
$(\sigma,n)$: $\{(0.2,125), (0.04,250), (0.1,500), (0.001,800) (0.02,1000), (0.01,2000)\}$. 
Each histogram was computed from $10^6$ ratios.
Dashed lines correspond to $\rho_{\tbox{PE}}(r)$ and $\rho_{\tbox{GOE}}(r)$, see 
Eqs.~(\ref{rhoPE},\ref{rhoGOE})
}
\label{Fig14}
\end{center}
\end{figure*}

Finally, it is relevant to add that once the scaling parameter of the quantities studied here was defined, 
it is expected that other properties related to the same quantities could also be scaled by the 
same scaling parameter. As an example, we validate the universality of the 
scaling parameter $\xi$ by applying it to $\rho(r)$, the probability distribution function of $r$.
In Fig.~\ref{Fig14} we present histograms of the probability distribution function of $r$, $\rho(r)$. 
Each panel displays six histograms for 
different combinations of $\sigma$ and $n$, while $\ell$ is tuned in order to produce the same 
value of $\xi$. Since the six histograms in each panel fall one on top of the other we can safely 
say that $\rho(r)$ is invariant for fixed $\xi$. 
In addition, we also include in each panel of Fig.~\ref{Fig14} the corresponding predictions for $\rho(r)$ for the PE 
and the GOE~\cite{ABG13}:
\begin{equation}
\label{rhoPE}
\rho_{\tbox{PE}}(r)=\frac{2}{(1+r)^2}
\end{equation}
and 
\begin{equation}
\label{rhoGOE}
\rho_{\tbox{GOE}}(r)=\frac{27}{4} \frac{r(1+r)}{(1+r+r^2)^{5/2}} ,
\end{equation}
respectively. Note that there is a perfect agreement of $\rho(r)$ 
with $\rho_{\tbox{PE}}(r)$ and $\rho_{\tbox{GOE}}(r)$ when $\xi=0.1$ and $\xi=3$, respectively. 
As expected, for $\xi=1$, i.e.~in the PE-to-GOE transition regime, we observe that the shape of 
$\rho(r)$ is in between the PE and the GOE predictions.

We hope that our work may motivate further analytical as well as numerical studies on
non-uniform random networks models and their applications to real-world systems.


\section*{Acknowledgements}
J.A.M.-B. acknowledges financial support from CONACyT (Grant No.~A1-S-22706) and
BUAP (Grant No.~100405811-VIEP2021).
E.E. thanks financial support from Ministerio de Ciencia, Innovacion y Universidades, Spain for the Grant 
No.~PID2019-107603GB-I00 ``Hubs-repelling/attracting Laplacian operators and related dynamics on graphs/networks".


\end{document}